# Literatura científica citada en patentes: Un indicador de Transferencia Tecnológica en las universidades portuguesas

Carmen Gálvez

**cgalvez@ugr.es**

Departamento de Información y Comunicación, Universidad de Granada, 18071, Granada, España. https://orcid.org/0000-0001-7454-1254

**Resumen:** El estudio tiene como objetivo identificar el proceso de transferencia de la ciencia a la tecnología que se produce en las principales universidades públicas portuguesas. La metodología se basó en el análisis de la literatura científica citada en patentes. Los datos se obtuvieron de la base de datos de patentes Lens. Se recuperaron 10.514 artículos científicos citados en patentes. Se realizó un análisis descriptivo de los datos. Se crearon mapas científicos para visualizar las principales tendencias de investigación. Los resultados mostraron una repercusión valiosa de la investigación académica en determinadas disciplinas científicas, como Química, Biología, Ciencias de los Materiales y Medicina. Los principales frentes de la investigación fueron el cáncer, las nanopartículas, los biomateriales, la ingeniería de tejidos o la biología molecular. En conclusión, la investigación producida por las universidades portuguesas ha generado conocimiento relevante para las invenciones patentadas y el flujo ciencia-tecnología dentro de sectores específicos.

## *Scientific literature cited in patents: A Technology Transfer indicator in Portuguese universities*

***Abstract:*** The study aims to identify the process of transfer from science to technology that occurs in the main Portuguese public universities. The methodology was based on the analysis of the scientific literature cited in patents. Data was obtained from the Lens patent database. 10,514 scientific articles cited in patents were retrieved. A descriptive analysis of the data was performed. Science maps were created to visualize the main research trends. The results showed a valuable impact of academic research in certain scientific disciplines, such as Chemistry, Biology, Materials Sciences and Medicine. The main research fronts were cancer, nanoparticles, biomaterials, tissue engineering or molecular biology. In conclusion, the research produced by Portuguese universities has generated relevant knowledge for patented inventions and the science-technology flow within specific areas.

***Keywords:*** Technology transfer; Research evaluation; Patents; Citations; Universities; Portugal.

## 1. Introducción

Las políticas de ciencia, tecnología e innovación constituyen un instrumento estratégico con el que cuenta los países para guiar la planificación y la toma de decisiones en la Investigación, el Desarrollo y la innovación (I+D+i). Las patentes juegan un papel esencial en la vinculación entre la investigación científica, la tecnología y la innovación. La transferencia de tecnología constituye el conjunto de actividades dirigidas a la difusión de conocimientos con el fin de facilitar su aplicación en los sectores productivos e industriales. En este sentido, las patentes indican una transferencia de conocimiento a la innovación industrial (Okubo, 1997; Velayos-Ortega & López-Carreño, 2023). Las patentes son documentos públicos que exponen las características técnicas de productos o procesos industriales, pero describen la invención no la innovación, ya que esta última supone una comercialización con éxito (Callon et al., 1995). A pesar de esta limitación, las patentes tienen una calidad excepcionalmente diferente (Silverberg & Verspagen, 2007), los datos de patentes constiuyen una fuente de información valiosa e inestimable, han sido muchos los intentos por proporcionar indicadores de su capacidad para la evaluación de la investigación, influir en los sectores productivos o identificar el flujo entre ciencia-tecnología.

Identificar el proceso de transferencia de conocimiento y tecnología a los sectores industriales es una tarea muy compleja en la que influyen diferentes factores (características de la oficina de patentes, políticas científicas de los países, características de las instituciones, de las empresas o de los diversos sectores tecnológicos). El análisis de patentes se ha empleado durante mucho tiempo como una herramienta analítica útil para el análisis de oportunidades tecnológicas (Yoon, 2010). Aunque el vínculo entre la actividad científica y la actividad empresarial está suficientemente respaldada en diversos trabajos (Etzkowitz et al., 1998; Leydesdorff & Etzkowitz, 1998; Rothaermel et al., 2007; Dias Daniel & Alves, 2020), los estudios sobre la dinámica detrás de tales procesos siguen siendo escasos (Callaert et al., 2015).

Durante un periodo de tiempo, las únicas medidas de patentes utilizadas en la investigación fueron el simple recuento del número de patentes asignadas a instituciones, empresas o países. Sin embargo, este indicador es considerado, por algunos, el más difícil de interpretar (Callon et al., 1995), debido a que dependen también de otros muchos factores (como los sectores tecnológicos, las compañías o las diversas legislaciones nacionales). En la actualidad, el indicador de impacto de patentes más utilizado es el número de citas recibidas, diversos estudios han establecido una correlación entre las citas de patentes y el valor de la patente (García-Escudero Márquez & López López, 1997; Harhoff et al., 1999; Jaffe et al., 2000; Harhoff et al., 2003; Zhang et al., 2017; Noruzi, 2022).

Una alternativa, para estudiar el proceso de transferencia de la ciencia a la tecnología, fue el análisis del papel que pueden desempeñar las citas de artículos científicos en las patentes. El planteamiento partió de la siguiente hipótesis: si para calcular el impacto de una publicación científica se utilizan habitualmente las citas recibidas en revistas científicas, de igual modo para calcular el impacto tecnológico de una publicación científica se podría utilizar las citas recibidas en las patentes (Callaert et al., 2006). No obstante, las citas o referencias en las patentes difieren de las citas científicas, distinguiendose dos tipos de documentos (Van Raan, 2017; Velayos-Ortega & López-Carreño, 2021):

- Referencias a otras patentes (LP, Literatura Patente). Su análisis permite conocer la identidad, ubicación y tipología de las empresas o entidades solicitantes de las patentes.
- Referencias a la literatura científica (LNP, Literatura No-Patente). su análisis permite obtener información acerca de los investigadores, instituciones, colaboraciones, revistas citadas, carácter básico, o aplicado, de la investigación, las temáticas o la proximidad en el tiempo entre la patente citante y la literatura citada.

El inicio del análisis métrico de las referencias de la literatura científica citada en documentos de patentes comenzó en USA (Narin & Olivastro, 1992). Estas investigaciones probaron que la ciencia, financiada con fondos públicos, constituye el pilar fundamental del avance industrial, demostrando que la ciencia pública se encuentra en el centro de la mayor parte de los procesos de innovación empresarial. Muchos estudios han reconocido que las citas de la literatura científica en las patentes son una métrica útil para medir el impacto de la tecnología y pronosticar actividades de innovación empresarial (Acosta Seró & Coronado Guerrero, 2002; Jaffe & Trajtenberg, 2002; Van Raan, 2017; Wang et al., 2019; Yamashita, 2020; Kim & Kim, 2021). En este contexo, el análisis de patentes constituye, en la actualidad, un campo en crecimiento que abarca el análisis de datos de patentes, el análisis de la literatura científica, la limpieza de datos, la minería de textos, el aprendizaje automático, el mapeo geográfico o la visualización de datos (Oldham & Kitsara, 2016). Este trabajo se encuadra en la línea de investigación del análisis métrico de la literatura científica citada en las invenciones patentadas.

## 2. Objetivos

El objetivo de este trabajo fue identificar la conexión entre la investigación académica y la transferencia de conocimiento, que se genera en las principales universidades portuguesas, a partir de las citaciones de literatura científica (LNP) en los documentos de patentes. Para describir el flujo entre ciencia-tecnología, se seleccionaron seis de las principales universidades públicas de Portugal: Universidad de Lisboa, Universidad de Oporto, Universidad de Coimbra, Universidad de Aveiro, Universidad de Miño y Universidad Nueva de Lisboa. Se excluyó el resto y no se tuvieron en cuenta otras instituciones de educación superior, es decir, institutos politécnicos y universidades privadas, porque solo se seleccionaron las universidades públicas portuguesas con mayor número de publicaciones científicas citadas en patentes. El diseño del estudio se dirigió a responder a las siguientes Preguntas de Investigación (PI):

PI-1 ¿Cuál es la evolución de la literatura científica citada en patentes?

PI-2 ¿Cuáles son los investigadores más productivos?

PI-3 ¿Cuáles son los Campos de Estudio y las instituciones más activas?

PI-4 ¿Cuáles son las revistas y editoriales más productivas?

PI-5 ¿Cuáles son las palabras-clave más frecuentes?

PI-6 ¿Cuáles son los principales frentes de investigación?

# 3. Metodología

El procedimiento utilizado se ha basado en el uso de métodos bibliométricos. Se han aplicado indicadores cuantitativos de producción e indicadores relacionales para detectar la estructura temática y conceptual del campo analizado. La metodología aplicada se ha desarrollado en las siguientes etapas (Börner et al., 2003; Noyons et al., 1999): 1) recopilación de los datos; 2) análisis de los datos; 3) visualización de los datos; y 4) interpretación de los datos. Para el procesamiento estadístico y descriptivo de los datos, así como para la visualización de datos se emplearon las herramientas de análisis y mapeo *RStudio Bibliometrix* versión 4.1.0 (Aria & Cuccurullo, 2017) y *VOSviewer* (Van Eck & Walkman, 2010).

## 3.1. Recopilación de datos

La fuente de información utilizada fue la base de datos de patentes Lens, caracterizada por la integración de metadatos procedentes de fuentes de acceso abierto de patentes, como *Espacenet*, *United States Patent and Trademark Office* (*USPTO*), *IP Australia*, *Patentscope* y de plataformas científico-académicas, como *CrossRef*, *ORCID*, *PubMed*, *Impactstory* y *CORE*. La base de datos Lens ofrece acceso a los metadatos de más de 225 millones de trabajos académicos y 127 millones de patentes. La ingeniería de Lens utiliza dos tipos principales de documentos: documentación de patentes y publicaciones académicas. La estrategia de búsqueda empleada consistió en seleccionar, dentro de los trabajos académicos (*Scholarly Works*), los trabajos citados por patentes (*Works Cited by Patents*), en concreto los artículos científicos citados en patentes. A continuación, se seleccionaron los campos: País, Institución y Tipo de publicación (Tabla 1).

| Base de datos | Período | Búsqueda | Resultados | Fecha búsqueda |
|---|---|---|---|---|
| Lens | 1972-2022 | Country = (*«Portugal»*) AND Institution = (*«University of Porto» OR «University of Lisbon» OR «University of Coimbra» OR «University of Aveiro» OR «University of Minho» OR «Universidade Nova de Lisboa»*) AND Publication Type = (*«journal article»*) | 10.514 | 30-06-2023 |

Tabla 1 – Estrategia de búsqueda en base de datos de patentes Lens

## 3.2. Análisis de los datos

Se realizó un análisis descriptivo de los datos extraídos de los artículos científicos citados en patentes. Para la identificación de la estructura conceptual y temática se utilizaron indicadores métricos multidimensionales a través de dos métodos:

1. Análisis de redes de palabras-clave. Este análisis consistió, primero, en generar una matriz cuadrada de P x P elementos (Palabras-clave por Palabras-clave), donde P es la palabras-clave a representar, a partir de las veces que aparecen las palabras-clave en los artículos científicos. Previamente se seleccionaron las palabras-clave con un mínimo de frecuencia de 5. A continuación, seleccionaron las 100 palabras-clave más representativas de la muestra analizada. El resultado fue una matriz de co-ocurrencias que reflejó el número de veces que un par de palabras-clave aparecieron conjuntamente en dos artículos científicos. A continuación, a la matriz se aplicó el índice de similitud Fuerza de Asociación (FA), o *Association Strength*, (Van Eck & Waltman, 2007), para normalizar los valores de co-ocurrencia. El índice FA se basa en la normalización de la intensidad de las asociaciones de las parejas de palabras-clave.
2. Análisis factorial. Se trata de un método estadístico multivariante cuyo propósito principal es definir la estructura subyacente de una matriz de datos. El análisis factorial es una técnica exploratoria multivariante de reducción de datos (como el Análisis de Correspondencias Simple, el Escalamiento Multidimensional o el Análisis de Correspondencias Múltiples), que sirve para encontrar grupos homogéneos de variables a partir de un conjunto de variables observadas. En este trabajo, se aplicó Análisis de Correspondencias Múltiples (MCA, *Multiple Correspondence Analysis*) (Greenacre & Blasius, 2006). Primero se creó una matriz A x P (Artículo por Palabras-clave), teniendo en cuenta el número de artículos donde aparecieron las palabras-clave. Previamente, se limitó a 100 el número de palabras-clave a representar. A continuación, se estableció un método para clasificar las palabras-clave en grupos que fueran lo más homogéneos posibles. La estrategia para definir los grupos se basó en el análisis de clúster k-medias (*k-means*), o análisis de conglomerados no jerárquicos. La función del análisis de clúster de k-medias fue maximizar la homogeneidad de ítems dentro de cada grupo. El número de clústeres se fijó en cuatro (*k=4*).

## 3.3. Visualización de los datos

El análisis de redes de palabras-clave y del análisis factorial se representaron en dos mapas bidimensionales (2D):

1. Mapa etiquetado. La selección de la red de palabras-clave se visualizó en un mapa, en el que las palabras-clave, en forma de nodos, fuertemente relacionadas se ubicaron lo más cerca posible y las palabras-clave débilmente relacionadas se ubicaron lejos. Para posicionar los nodos se utilizó la técnica Visualización de Similaridades (VOS, *Visualization of Similarities*) (Van Eck & Waltman, 2010). La técnica de mapeo VOS permitió ejecutar diferentes algoritmos de agrupamiento, o *clustering*, para posicionar y clasificar las palabras-clave en grupos similares, equiparables a tendencias de investigación. La red de co-ocurrencia de palabras-clave se representaron por un círculo y se identificaron por una etiqueta. Cuanto mayor fue el peso de un nodo, mayor fue el círculo y la etiqueta. El color aleatorio de los nodos estuvo determinado por el grupo al que pertenece cada palabra-clave. Las líneas entre los elementos representaron los enlaces o vínculos.
2. Mapa factorial. En esta representación las palabras-clave similares se distribuyeron lo más cerca en las zonas del mapa. En la representación del mapa se tuvo en cuenta las posiciones relativas de los puntos de palabras-clave, su distribución a lo largo de las dimensiones y cómo los artículos se agruparon juntos en el mapa factorial. Con la aplicación de este método se obtuvo la estructura temática de la literatura científica citada en patentes, que se interpretó como los principales focos de investigación vinculados a los artículos científicos más relevantes.

## 4. Resultados y Discusión

Se obtuvo un total de 10.514 artículos de investigación que recibieron citas en los documentos de patentes. Los resultados nos permitieron identificar los datos generales sobre los trabajos científicos que sirvieron para apoyar a una o más patentes (Tabla 2).

| Descripción | Resultados |
|---|---|
| *Artículos citados en patentes* | 10.514 |
| *Periodo* | 1972-2022 |
| *Revistas* | 2.520 |
| *Autores* | 45.416 |
| *Autores de artículos de un solo autor* | 112 |
| *Palabras-clave* | 7.797 |

Tabla 2 – Análisis descriptivo: Datos de la colección

La evolución anual de los artículos citados en las patentes evidenció un aumento progresivo a partir de la década de 1990 (Figura 1). Se constató que el intervalo de tiempo que transcurrió entre la publicación de los artículos y las citas en las patentes fue más largo que en otro tipo de citas científicas. Lo anterior explica que disminuyera el número de citas a partir de 2020 (es decir, se requiere un lapso de tiempo mayor para que la literatura científica sea citada en las patentes).

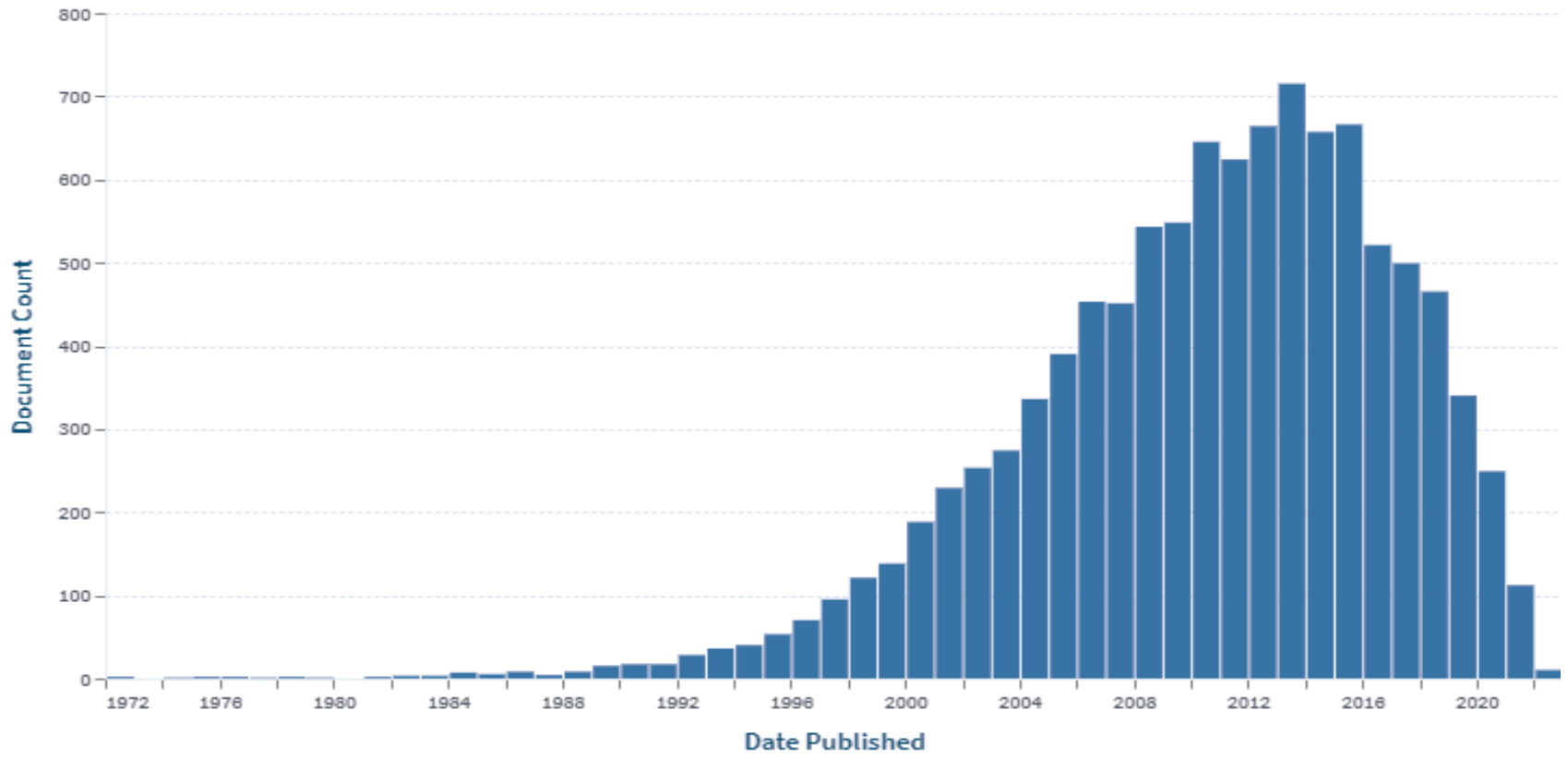

Figura 1 – Número de documentos y evolución anual de los artículos científicos citados en patentes (1972-2022)

### 4.1. Análisis descriptivo de los autores más productivos

Los resultados mostraron que, entre los autores más citados en patentes, dentro de los 45.416 autores totales detectados, se encontraron (Tabla 3): Rui L. Reis (Departamento de Ingeniería de Polímeros, Universidad de Miño), João F. Mano (Departamento de Química, Universidad de Aveiro), Alirio E. Rodrigues (Departamento de Ingeniería Química, Universidad de Oporto), José A. Teixeira (IBB-Instituto de Biotecnología y Bioingeniería, Centro de Ingeniería Biológica, Universidad de Miño), Artur M. S. Silva (Departamento de Química, Universidad de Aveiro), Elvira Fortunato (Departamento de Ciencia de los Materiales, Universidad Nueva de Lisboa), João Rocha (Departamento de Química, Universidad de Aveiro), Armando J. D. Silvestre (Departamento de Química, Universidad de Aveiro), Rodrigo Martins (Departamento de Ciencia de los Materiales, Universidad Nueva de Lisboa), Artur Cavaco-Paulo (Departamento de Ingeniería Biológica, Universidad de Miño).

| Rank | Autor | Institución | Nº de Artículos citados en patentes |
|---|---|---|---|
| 1 | *Reis, Rui L.* | Universidad de Miño | 286 |
| 2 | *Mano, J. F.* | Universidad de Aveiro | 156 |
| 3 | *Rodrigues, A. E.* | Universidad de Oporto | 142 |
| 4 | *Teixeira, J. A.* | Universidad de Miño | 117 |
| 5 | *Silva, A. M. S.* | Universidad de Aveiro | 88 |
| 6 | *Fortunato, E.* | Universidad Nueva de Lisboa | 79 |
| 7 | *Rocha, J.* | Universidad de Aveiro | 73 |
| 8 | *Silvestre, A. J. D.* | Universidad de Aveiro | 72 |
| 9 | *Martins, R.* | Universidad Nueva de Lisboa | 71 |
| 10 | *Cavaco-Paulo, A.* | Universidad de Miño | 67 |

Tabla 3 – Análisis descriptivo de autores: Top-10 autores más productivos

### 4.2. Análisis descriptivo de los principales Campos de Estudio

| Rank | Campos de Estudio | Nº de Artículos citados en patentes |
|---|---|---|
| 1 | Química | 3.374 |
| 2 | Biología | 2.645 |
| 3 | Ciencia de los Materiales | 1.934 |
| 4 | Medicina | 1.040 |
| 5 | Bioquímica | 953 |
| 6 | Ciencias de la Computación | 688 |
| 7 | Química Orgánica | 654 |
| 8 | Ingeniería Química | 626 |
| 9 | Medicina Interna | 602 |
| 10 | Cromatografía | 572 |

Tabla 4 – Análisis descriptivo de Campos de Estudio: Top-10 disciplinas más activas

Los resultados mostraron los Campos de Estudio más activos (Tabla 4). Como ya se ha mencionado, la literatura científica citada en patentes constituye un puente en el flujo ciencia-teconología. Las disciplinas de Química y Biología, seguida de Ciencias de los Materiales y Medicina, fueron las que recibieron con mayor intensidad científica en esa interconexión.

### 4.3. Análisis descriptivo de las universidades más activas según el Campo de Estudio

Los resultados mostraron los Campos de Estudio de las instituciones más activas (Tabla 5), destacando la Universidad de Oporto en las disciplinas de Química, Biología, Medicina y Bioquímica. La Universidad de Aveiro destacó en el campo de las Ciencias de los Materiales.

| Campo de Estudio | Universidad | Nº de Artículos citados en patentes |
|---|---|---|
| *Química* | Universidad de Oporto<br>Universidad de Coimbra<br>Universidad de Aveiro<br>Universidad de Miño | 1.059<br>602<br>596<br>534 |
| *Biología* | Universidad de Oporto<br>Universidad de Coimbra<br>Universidad Nueva de Lisboa<br>Universidad de Lisboa | 970<br>486<br>474<br>470 |
| *Ciencias de los Materiales* | Universidad de Aveiro<br>Universidad de Miño<br>Universidad de Oporto<br>Universidad de Coimbra | 615<br>575<br>415<br>219 |
| *Medicina* | Universidad de Oporto<br>Universidad de Coimbra<br>Universidad de Lisboa<br>Universidad de Miño | 464<br>252<br>244<br>54 |
| *Bioquímica* | Universidad de Oporto<br>Universidad Nueva de Lisboa<br>Universidad de Miño<br>Universidad de Lisboa | 308<br>197<br>173<br>161 |

Tabla 5 – Análisis descriptivo de las universidades más activas

### 4.4. Análisis descriptivo de las revistas y editoriales más productivas

Las revistas más relevantes, dentro del total de las 2.520 revistas identificadas, con mayor número de artículos científicos, fueron publicaciones en editoriales académicas de prestigio internacional como (Tabla 6): *Biomaterials* (Elsevier), *Scientific Reports* (Springer Science), *Journal of Agricultural and Food* Chemistry (American Chemical Society) o *Biotechnology and Bioengineering* (Wiley).

| **Editorial académica** | **Título de la Revista** | **Nº de Artículos citados en patentes** |
|---|---|---|
| Elsevier | *Biomaterials*<br>*Acta Biomaterialia*<br>*Thin Solid Films* | 59<br>45<br>43 |
| Springer Science | *Scientific Reports*<br>*Journal of Materials Science*<br>*Nature* | 65<br>33<br>31 |
| American Chemical Society | *Journal of Agricultural and Food Chemistry*<br>*Industrial & Engineering Chemistry Research*<br>*Biomacromolecules* | 57<br>41<br>28 |
| Wiley | *Biotechnology and Bioengineering*<br>*European Journal of Organic Chemistry*<br>*FEBS Letters* | 31<br>20<br>18 |

Tabla 6 – Análisis descriptivo de las revistas y editoriales más productivas

## 4.5. Análisis descriptivo de las principales palabras-clave

Entre las palabras-clave más frecuentes, dentro del total de las 7.797 palabras-clave detectadas, se encontraron (Tabla 7): "*Nanoparticles*", "*Cancer*", "*Drug Delivery*" o "*Biomarkers*".

| ***Keywords*** | **Nº de frecuencias** | ***Keywords*** | **Nº de frecuencias** |
|---|---|---|---|
| *Nanoparticles* | 48 | *Apoptosis* | 18 |
| *Cancer* | 40 | *Biomaterials* | 17 |
| *Drug Delivery* | 39 | *Breast Cancer* | 17 |
| *Biomarkers* | 38 | *Ionic Liquids* | 17 |
| *Alzheimer's Disease* | 30 | *Neurodegeneration* | 16 |
| *Parkinson's Disease* | 28 | *Aging* | 15 |
| *Chitosan* | 26 | *Antimicrobial* | 14 |
| *Tissue Engineering* | 25 | *Antioxidant Activity* | 14 |
| *Inflammation* | 23 | *Mitochondria* | 14 |
| *Cytotoxicity* | 22 | *Oxidative Stress* | 14 |

Tabla 7 – Análisis descriptivo de las palabras-clave: Top-20 términos más frecuentes

## 4.6. Visualización de la estructura conceptual y temática

El resultado del análisis de la red de palabras-clave fue un mapa etiquetado 2D, en el que se identificaron tres grandes focos de interés citadas. Las palabras-clave dentro de un mismo grupo (G) representaron las diferentes tendencias de investigación (Figura 2):

- G1. Agrupó palabras-clave como "*cancer*", "*inflammation*", "*biomarkers*", "*breast cancer*", "*diagnosis*", "*gold nanoparticles*", "*cytotoxicity*", "*angiogenesis*",

"*exosomes*", "*extracellular vesicles*", "*cancer therapy*", "*dna methylation*", "*immune response*", "*antioxidant activity*", "*microRNAs*" o "*gene expression*". Este aglomerado se vinculó con las investigaciones relacionadas las causas, el diagnóstico y el tratamiento del cáncer. La red de palabras-clave reflejó las innovaciones sobre la terapia celular personalizada y la terapia dirigida para tratar el cáncer con células inmunitarias del propio paciente. Los científicos cada vez entienden mejor los cambios en el ADN y ARN, esto ha dado lugar a los avances recientes en los tratamientos de enfermedades, que se mostró como una tendencia de investigación.

- G2. Agrupó palabras-clave como "*nanoparticles*", "*drug delivery*", "*tissue engineering*", "biomaterials", "drug delivery systems", "*hidrogel*", "*silk fibroin*", "*hydrogels*", "*biocompatibility*", "*gene therapy*" o "*polymers*". Esta agrupación se relacionó con la tendencia de investigación de la nano-ingeniería de tejidos, dirigida a desarrollar sustitutos para los órganos o los tejidos dañados. Este foco se investigación también mostró el desarrollo de nanopartículas y su expansión a una amplia gama de aplicaciones clínicas.
- G3. En este clúster aparecieron palabras-clave como "*parkinson's disease*", "*alzheimer's disease*", "*neurodegeneration*", "*aging*", "*clinical trials*", "*biosensors*", "*mesenchymal stem cells*", "*nanotechnology*", "*neuroprotection*", "*clinical trials*", "*glioblastoma*", "*amyotrophic lateral sclerosis*" o "*neuroinflammation*". Este frente extraordinario de investigación reflejó las grandes posibilidades que ofrecen los avances en genética molecular, biología molecular y medicina genómica para afrontar el estudio de las enfermedades neurodegenerativas.

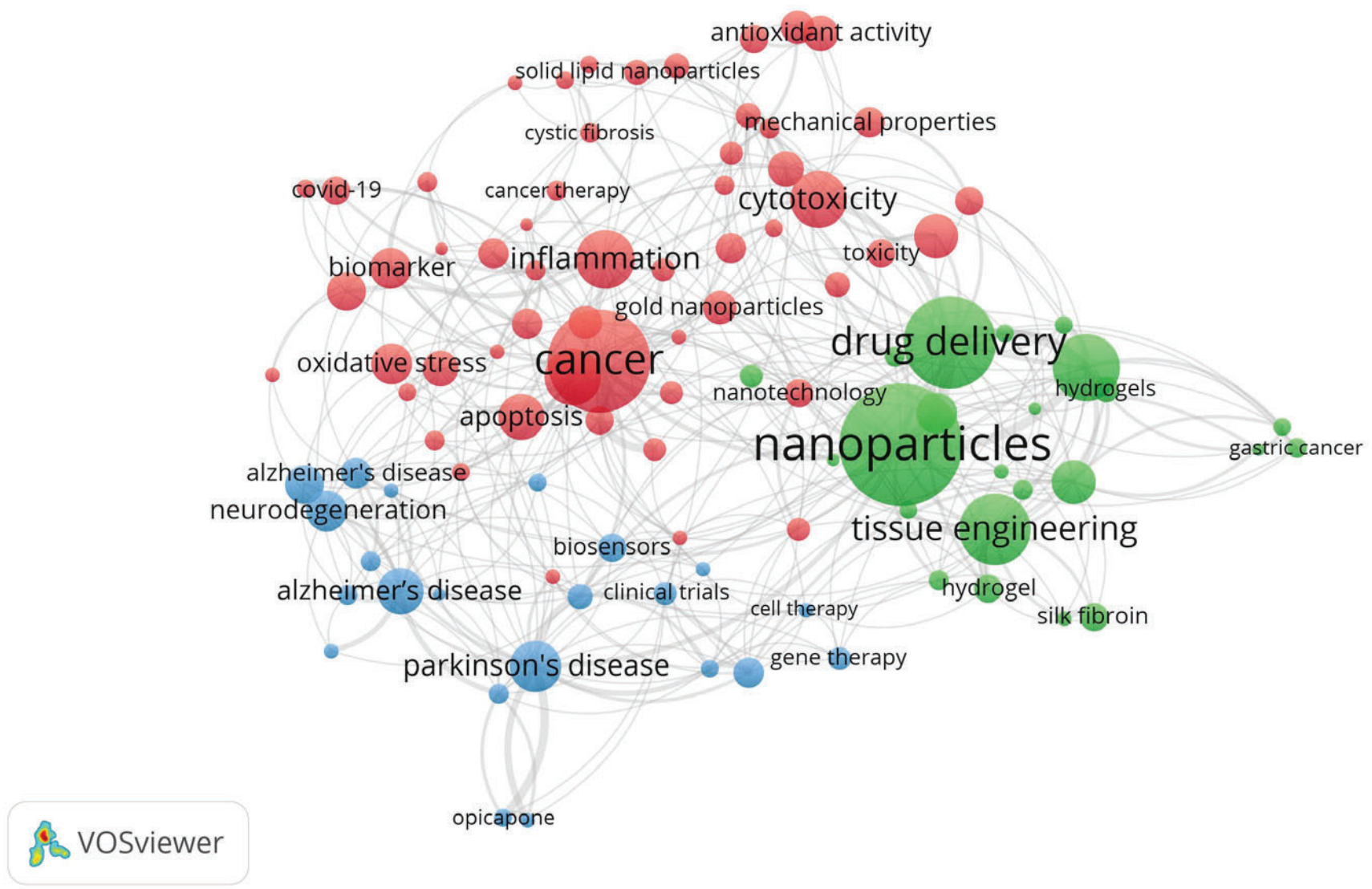

Figura 2 – Visualización del mapa etiquetado: Red de palabras-clave

En cuanto al análisis factorial, el mapa 2D resultante presentó cuatro clústeres, que representaron las principales áreas de influencia científica. El tamaño de los puntos, dentro de cada clúster, fue proporcional a las contribuciones de los artículos científicos. La proximidad entre palabras-clave correspondió a las materias compartidas. Las palabras-clave que se ubicaron cerca, las unas de las otras, se debió a que una gran proporción de artículos las trataron conjuntamente. La combinación de las dos dimensiones del mapa también proporcionó información interesante sobre las relaciones entre las palabras-clave, reflejando los polos característicos de las diferentes orientaciones temáticas. Los diferentes cuadrantes del mapa se interpretaron de la siguiente forma (Figura 3):

- Parte central. El núcleo del mapa representó la posición promedio de todos los artículos que conformaron el centro de interés del campo de investigación, relacionado con la terapia celular para tratar el cáncer, con palabras-clave como "*cancer*", "*gold nanoparticles*" o "*biomarker*".
- Parte superior izquierda. En el lado superior del mapa se identificaron artículos relevantes, por su posición cerca del centro, vinculados a los nuevos tratamientos de enfermedades que incorporan nanopartículas y células madre, con palabra-clave clave como "*nanoparticles*", "*chitosan*", "*mesenchymal.stem.cells*" o "*neuroprotection*".
- Parte inferior izquierda. En esta región del mapa se ubicaron artículos vinculados a la actividad antioxidante y antimicrobiana como sustancias químicas para proteger el deterioro de las células y el desarrollo de enfermedades, con palabras-clave como "*antioxidant.activity*", "*antimicrobial.activity*" o "*phenolic.compounds*".

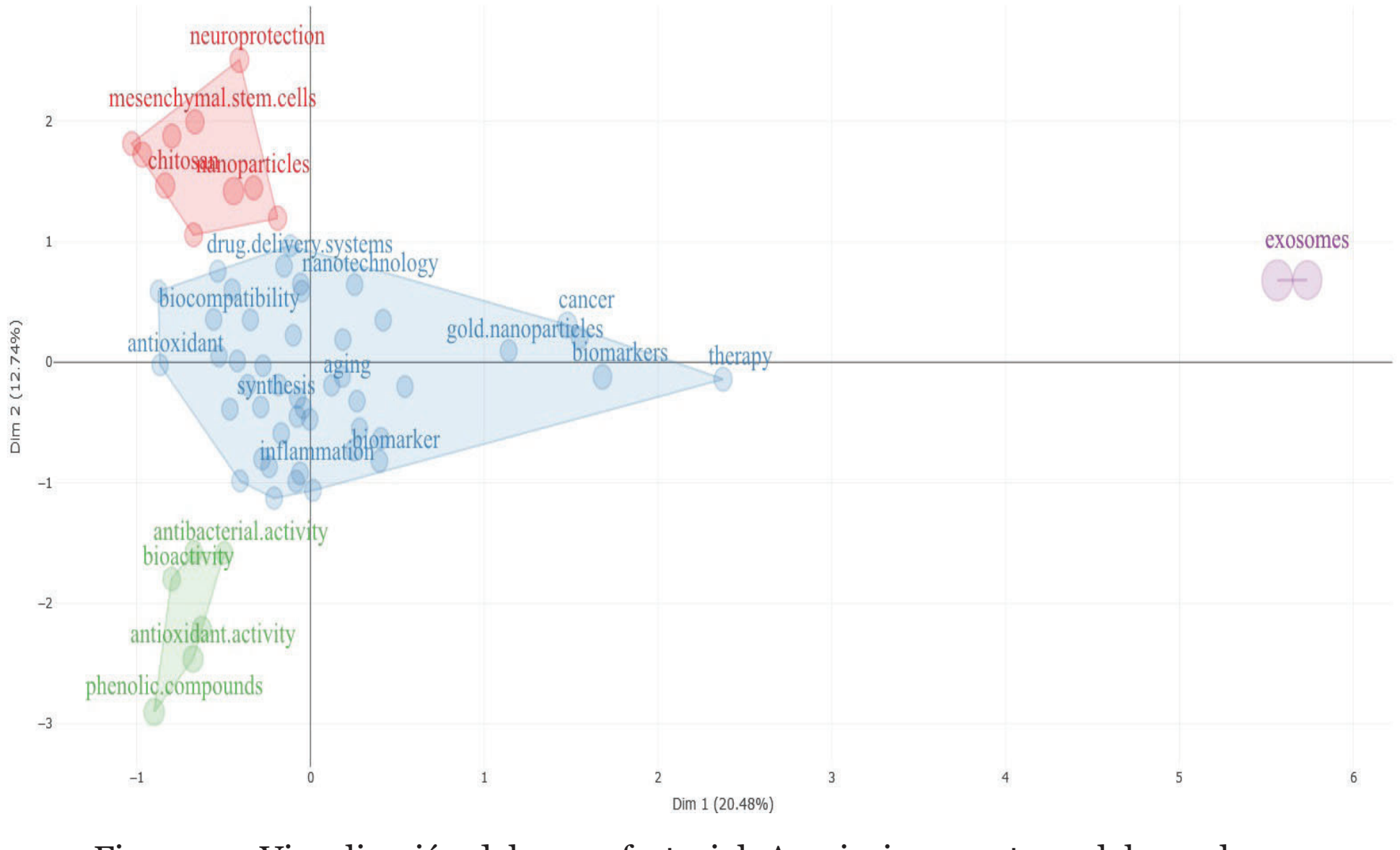

Figura 3 – Visualización del mapa factorial: Asociaciones entre palabras-clave y artículos citados en patentes

- Parte superior derecha. En esta parte del mapa se mostró el incremento de estudios y publicaciones sobre los exosomas, nanopartículas lípidas relacionadas con la metástasis del cáncer. La ubicación en el mapa de este frente de interés científico, con la palabra-clave "*exodomes*" evidenció que se trata de un campo de investigación emergente. Se constató el aumento de publicaciones que estudian la biogénesis y los efectos biológicos de los exosomas.

## 5. Conclusiones

La literatura científica citada en patentes se ha proyectado como un indicador clave para analizar la transferencia de conocimiento desde el ámbito académico a la tecnología, esto es, en el flujo ciencia-tecnología. En este trabajo se han analizado los artículos científicos citados en patentes, generados por investigadores pertenecientes a las principales universidades portuguesas (Universidad de Lisboa, Universidad de Oporto, Universidad de Coímbra, Universidad de Aveiro, Universidad de Miño y Universidad Nueva de Lisboa). El estudio mostró que la evolución anual de los artículos citados en las patentes aumentó progresivamente a partir de la década de 1990, se constató que se requiere un intervalo de tiempo mayor para que la literatura científica sea citada en las patentes. Se identificaron los investigadores más productivos, destacando, entre otros: Rui L. Reis (cuya investigación se centra en la ingeniería de tejidos, medicina regenerativa, biomateriales, biomimética, células madre y polímeros biodegradables), João F. Mano (cuya investigación se dirige a la ingeniería de tejidos, medicina regenerativa y células madre) o Alírio E. Rodrigues (cuyos intereses de investigación se situan en los campos de la ingeniería química, bioingeniería e ingeniería de materiales). Se detectó que los Campos de Estudio más activo fueron las disciplinas científicas de Química, Biología, Ciencia de los Materiales, Medicina y Bioquímica. Se reveló que la Universidad de Oporto destacó en las disciplinas de Química, Biología, Medicina y Bioquímica y que la Universidad de Aveiro destacó en el campo de las Ciencias de los Materiales. Se observó que las revistas citadas, más relevantes, fueron publicaciones de prestigio internacional (como *Biomaterials*, *Scientific Reports*, *Journal of Agricultural and Food* Chemistry, *Biotechnology and Bioengineering*). Se identificaron las principales editoriales académicas donde publicaron los investigadores, destacando Elsevier, Springer Science, American Chemical Society y Wiley. Se examinaron las principales palabras-clave utilizadas en los artículos científicos, predominando términos como "*Nanoparticles*", "*Cancer*", "*Drug Delivery*", "*Biomarkers*", "*Alzheimer's Disease*", "*Parkinson's Disease*" y "*Tissue Engineering*". La visualización de los mapas 2D mostraron la estructura conceptual y temática vinculada con las principales tendencias de investigación, relacionadas con focos emergentes de interés, como las aplicaciones clínicas de las nanopartículas, el cáncer, la ingeniería de tejidos, la genética molecular, la biología molecular, los exosomas o la medicina genómica. Reafirmando la premisa, según la cual, los campos de estudio en rápido desarrollo generalmente se basan más en el conocimiento científico reciente que los campos maduros, o campos que se encuentran en un estado de producción decreciente (Van Raan, 2017). En síntesis, los hallazgos identificados demostraron que los trabajos, procedentes de las principales universidades portuguesas, han tenido una repercusión relevante tanto en la investigación básica, generando conocimiento original, que ha influido en las invenciones patentadas, como en la investigación aplicada, en áreas específicas como la nanotecnología, las nanopartículas, los biomateriales, la bioingeniería o la nanomedicina.

Por último, el análisis de la literatura científica citada en patentes ofreció una aproximación valiosa pero parcial, debido a que las citas a la literatura científicas no aparecen en las patentes de todos los sectores tecnológicos por igual. Por tanto, la primera limitación de este estudio fue que hay patentes que no contienen citas a la literatura científica, y eso no implica que no se puedan vincular con la interconexión ciencia-tecnología. Otra limitación se presentó en las referencias científica que se extrajeron de la base de datos de patentes Lens, a pesar de que esta plataforma fue la que más prestaciones ofreció, en cuanto al análisis de las citas de publicaciones, también presentó algunas deficiencias con respecto a la normalización de datos. Sin embargo, a pesar de las limitaciones expuestas, la metodología emplea ha dado respuesta a las cuestiones inicialmente planteadas. En próximas líneas de investigación se pretende ampliar las instituciones analizadas, examinar el comportamiento de las citaciones de la literatura según los distintos sectores tecnológicos y mejorar el sistema de depuración de datos.

## Referencias